\documentclass[final]{aipproc}            % use final for the camera ready runs
\layoutstyle{8x11double}

\begin{document}

\title{Multi-wavelength Observations of LS I +61 303 with VERITAS, Swift and RXTE: 2006-2008}


\author{A.W. Smith}{
  address={Argonne National Laboratories, Argonne National Laboratory, 9700 S. Cass Avenue, Argonne, IL 60439, USA (awsmith@anl.gov)}
}

\author{the VERITAS collaboration}{
  address={http://veritas.sao.arizona.edu}
}

\begin{abstract}
A long term, multi-wavelength monitoring campaign on the TeV binary LS I +61 303 has been performed utilizing $>$300 GeV observations with VERITAS along with monitoring  in the 0.2-10 keV band by RXTE and Swift between September 2006 and February 2008. The source was detected by VERITAS as a variable TeV source with flux values ranging from 5-20$\%$ of the Crab Nebula flux with the strongest flux levels appearing around apastron. X-ray observations by RXTE and Swift show the source as a highly variable hard X-ray source with flux values varying in the range of 0.5-3$\times$10$^{-11}$ ergs cm$^{-2}$s$^{-1}$ over a single orbital cycle. The 2007-2008 RXTE data set also shows the presence of several extremely large flaring episodes presenting a flux of up to 7.2$\times$10$^{-11}$ ergs cm$^{-2}$s$^{-1}$, the largest such flare recorded from this source. Comparison of the contemporaneous TeV and X-ray data does not show a correlation at this time, however, the sparsity of data sets do not preclude the existence of such a correlation. 

\end{abstract}

\maketitle

%%%%%%%%%%%%%%%%%%%%%%%%%%%%%%%%%%%%%%%%%%%%
%% MAINMATTER
%%%%%%%%%%%%%%%%%%%%%%%%%%%%%%%%%%%%%%%%%%%%

\section{Background}

The galactic binary LS I +61 303 is one of the most heavily studied binary star systems in the Milky Way, and although the subject of many observational campaigns, the fundamental identification of the system (i.e. microquasar or binary pulsar) remains relatively unclear. From observations detailed in \cite{HC1981,C2005} it is clear that the system can be classified as a high mass X-ray binary (HMXB) located at a distance of 2 kpc; the components of the system consisting of a compact object in a 26.496($\pm$0.0028) day orbit around a massive BO Ve main sequence star. Periastron passage of the compact object occurs at $\phi=$0.23 (here $\phi$ represents the orbital phase ranging from 0.0 to 1, $\phi$=0.0 set at  JD 244336.775), with apastron passage occuring at 0.73, and inferior and superior conjunctions occuring at 0.26 and 0.16 respectively \cite{C2005}.  LS I +61 303 has been historically an object of interest due to its periodic outbursts at radio \cite{Par1998,Greg2002} and  X-ray energies \cite{L1997,Tay1996,GR2001,Har2000}. LS I +61 303 has also been identified with the EGRET source 3EG J0241+6103, a source which shows evidence for 26.5 modulation in the GeV band \cite{M2004}. More recently, LS I +61 303 has been detected as a variable TeV gamma-ray source \cite{Alb2006,Acc2008} with high emission appearing near apastron passage.

The two main competing scenarios to explain the system can be summarized as $\textit{microquasar}$ (i.e. non-thermal emission powered by accretion and jet ejection) or $\textit{binary pulsar}$ (i.e. non-thermal emission powered by the interaction between the stellar and pulsar winds). The microquasar model used to describe this system has several drawbacks (for example, lack of blackbody X-ray spectra), however the scenario has not been ruled out, for example see Romero et al. (2007). The binary pulsar model is possibly most strongly supported by the result of Dhawan (2006) in which VLBA monitoring of the system revealed a cometary radio structure around LS I +61 303 which was interpretated as the interaction between the pulsar and Be star wind structures. However, there is currently no evidence for the presence of a pulsar within the system either in radio or X-rays.

\section{Observations}
The VERITAS array of telescopes located in Southern Arizona (1268 m a.s.l, 31$^{\circ}$40'30''N, 110$^{\circ}$57'07'' W, \cite{W2002}) began full, 4 telescope array observations in April 2007 \cite{Gernot2007} and represents the most sensitive IACT array in the Northen Hemisphere. The array is comprised of four 12m diameter Davies-Cotton telescopes with tesselated mirror structures of 350 24m focal length hexagonal mirror facets (comprising a total mirror area of 100 m$^{2}$). Each telescope focuses light onto 499 pixel PMT cameras with a total field of view of 3.5$^{\circ}$. VERITAS is sensitive in the 100 GeV to 30 TeV energy regime with an energy resolution of $\sim$15-20$\%$ (energy and zenith angle dependent) and an angular resolution of $<0.1^{\circ}$ on an event by event basis. VERITAS observed LS I +61 303 from September 2006 until February 2008 resulting in a total of 58 hours of observations, covering nine 26.5 day orbital cycles. The observations were taken over a range of elevations from 52$^{\circ}$ to 61$^{\circ}$ with different combinations of 2,3, and 4 telescopes, resulting in an overall energy threshold of 300 GeV. See \cite{Acc2008} for a detailed description of the VERITAS data reduction and analysis procedures.
\begin{figure}
  \includegraphics[height=.25\textheight,width=0.35\textheight]{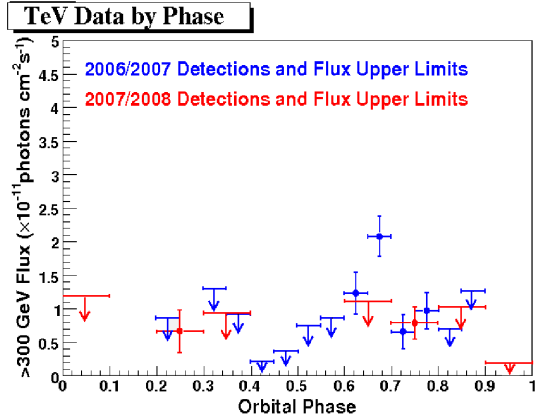}
  \caption{Detections and 99$\%$ flux upper limits resulting from the 2006-2008 VERITAS observations on LS I +61 303, binned by orbital phase.}
\end{figure}

The two RXTE \cite{Swank2004} data sets used for this work were accumulated first as a result of a Target of Opportunity observation request in October 2006, and then as a successful Cycle 12 Observation proposal for observations in 2007 and 2008. After quality selection, the October 2006 ToO request resulted in 9 ks of total observation time between 10/31/06 to 10/31/06. The observations, 1 ks pointings taken every other day, spanned the orbital phase bins of $\phi = 0.14 \rightarrow 0.83$. Observations were approved in 2007/8 for 1 ks pointing every other day from August 28, 2007 until February 2, 2008. These observations covered 6 full 26.496 day orbital cycles.  Both datasets were reduced by fitting each observation's spectra with an absorbed power-law between 3-10 kev and then extracting an integrated flux over this energy range. For details of the reduction and analyses procedures used please see \cite{Smith2008}.

The Swift-XRT observations span the period of September 2006 through September of 2007, with a total of 97.3ks observing time. For a description of Swift and the XRT instrument see \cite{G2004}. These data are comprised of many $\sim$1 ksec pointings, which are binned together over bin widths of approximately 1 day. The maximum span of a single binned observation is two days. The Swift-XRT data were screened and processed using the most recent versions of standard Swift tools: Swift Software version 2.8, ftools version 6.5, and XSPEC version 12.4.0. The data analysis procedure is similar to that described in \cite{H2007}.

\section{Results}

\subsection{TeV}

The data from the 2006/7 observing season comprised a total of 45.9 hours of observations (after data quality selection) during the orbital phases of 0.2$\rightarrow$0.9 with significant coverage of phases 0.4 to 0.9 (see Acciari et al. 2008). The source was detected at the 8.4$\sigma$ significance level for emission above 300 GeV. However, this detection only occurred between orbital cycles 0.5$\rightarrow$0.8 with the greatest TeV flux occuring between phases 0.6 and 0.8 at a peak flux of 10$\rightarrow$20$\%$ of the Crab Nebula flux (see figure 1). The source was able to be significantly detected during a single nights observations at times, most notably on the night of October 27, 2006 (MJD 54035, orbital phase 0.72), during which the source presented a 22.95$\%$ ($\pm4.3\%$) Crab Nebula flux at a significance of 6.1$\sigma$ (see figure 3). The differential energy spectrum extracted from the 2006/2007 observations (phases 0.5-0.8 only) is well fit by a powerlaw described by (2.39$\pm$0.32$_{stat}\pm$0.6$_{sys}$) $\times($E$^{-2.4\pm0.19_{stat}\pm0.2_{sys}}$)$\times$10$^{-12}$cm$^{-2}$s$^{-1}$TeV$^{-1}$. 

The 2007/8 season dataset is comprised of a total of 12 hours of four telescope observations taken between October 2007 and January 2008 spanning 4 separate orbital cycles. These observations had significant coverage occurring between 0.2$\rightarrow$0.4 and 0.6$\rightarrow$0.9. The 2007-8 observations show a much less active source, with a detection at a significance at the 4$\sigma$ level for emission above 300 GeV. The source was not significantly detected during the phases of 0.4$\rightarrow$0.7 as in previous observing seasons \cite{Alb2006,Acc2008},  with a significant detection resulting from  observations during the orbital phases of 0.7$\rightarrow$0.8 only. However, this variation in detection may be due to uneven orbital phase sampling more than intrinsic source variation. The detection was not significant enough to produce an energy spectrum. It should be noted that the 2007-8 season of observations included a considerable amount of moonlight observations (which are not presented here) that raise the significance of the detection of LS I +61 303. The results from these observations will be presented in a forthcoming publication \cite{Accb2008}.

\begin{figure}
  \includegraphics[height=.25\textheight,width=0.35\textheight]{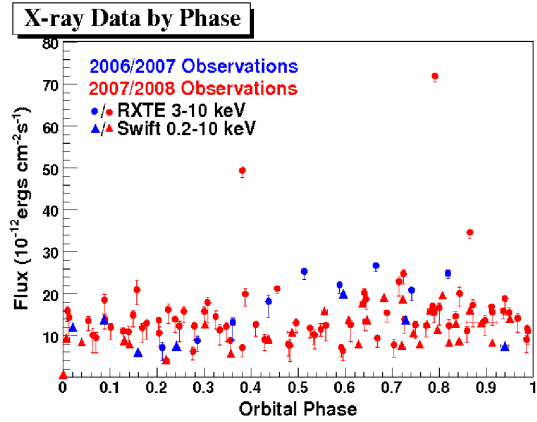}
  \caption{The 2006-2008 X-ray monitoring of LS I +61 303 performed by RXTE and Swift binned by orbital phase.}
\end{figure}
\begin{figure}
  \includegraphics[height=.45\textheight,width=\textwidth]{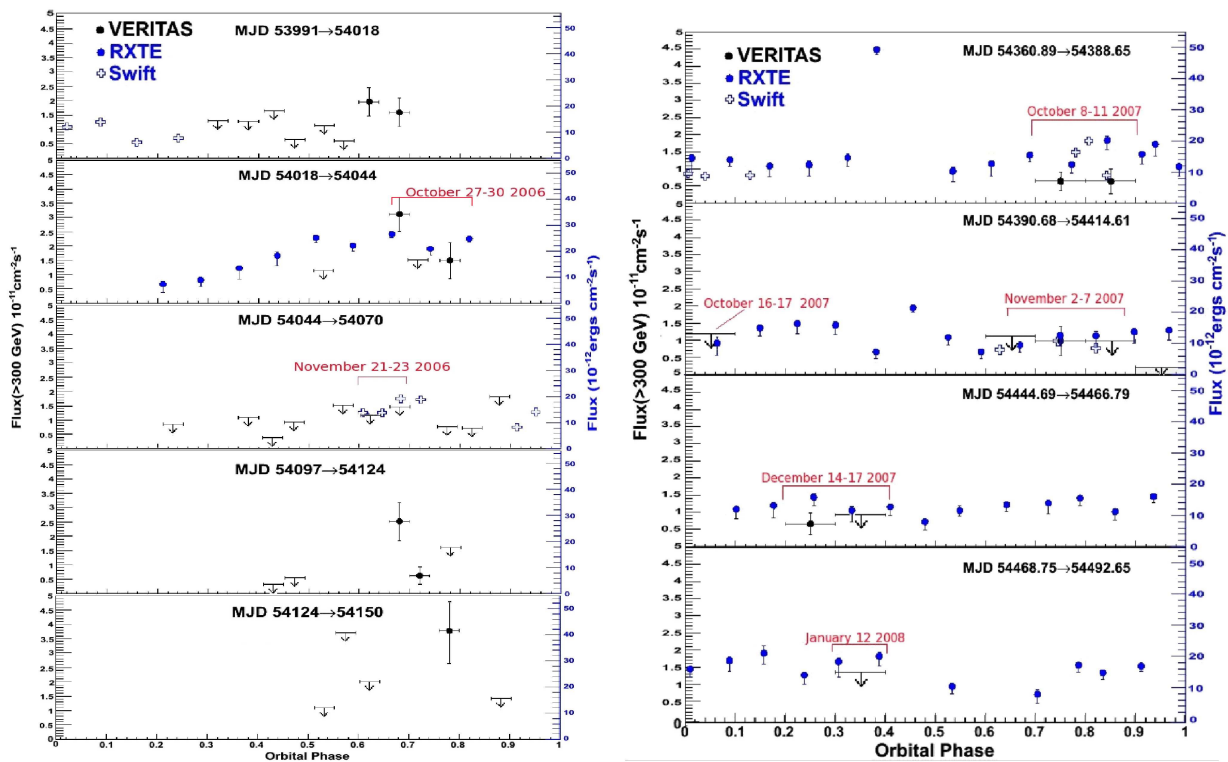}
  \caption{The overlapping TeV and X-ray from the 2006-8 multiwavelength campaign.}
\end{figure}
\begin{figure}
  \includegraphics[height=.25\textheight,width=0.35\textheight]{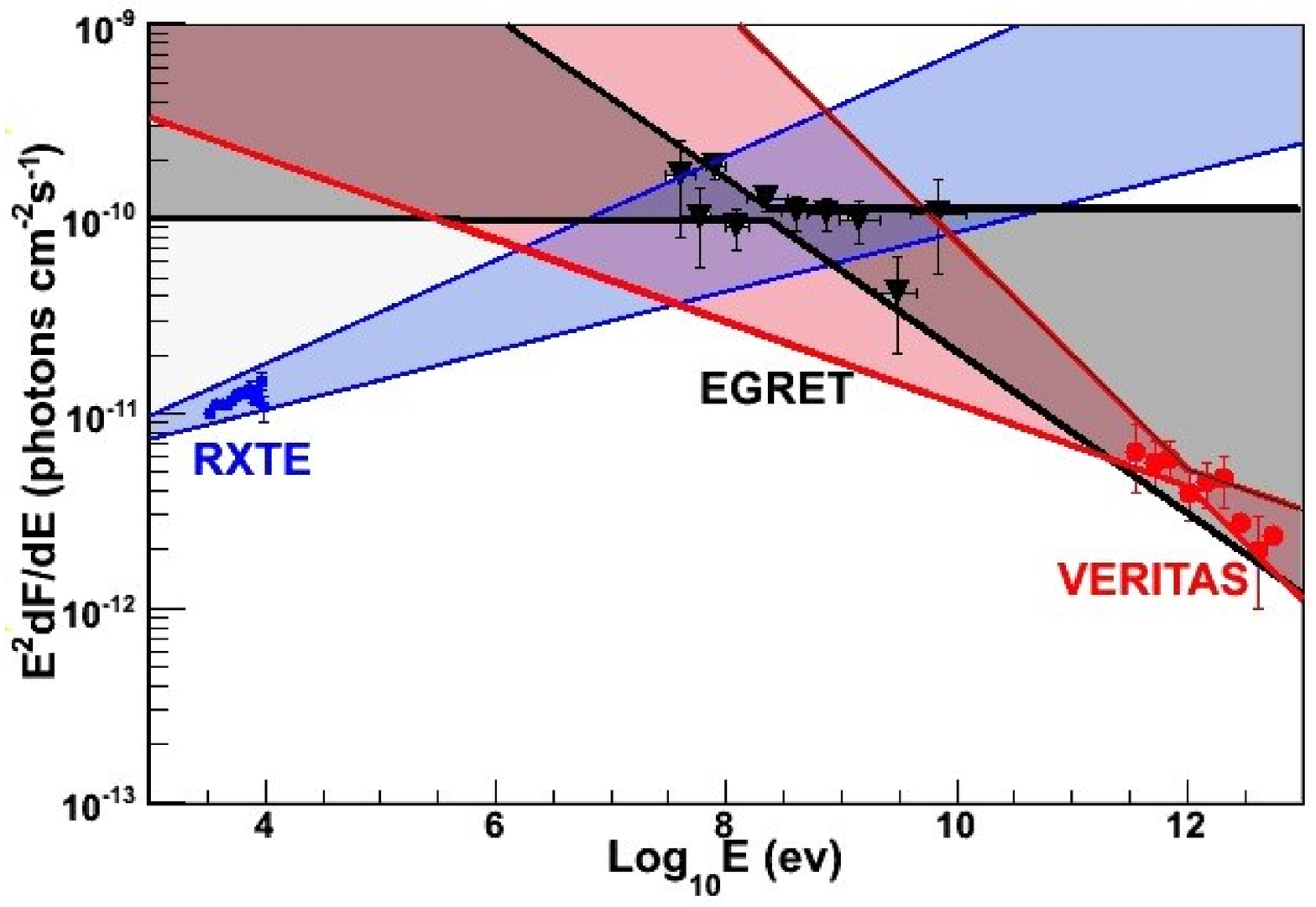}
  \caption{The X-ray, GeV, and TeV spectra described in the text along with the ranges allowed by extrapolating the associated errors in each band. }
\end{figure}
\subsection{X-ray}
Shown in figure 2 are the results of the 2006-8 X-ray observations conducted by RXTE and Swift. As can be seen, the X-ray flux in the 2006-7 season as seen by both instruments shows a highly variable flux oscillating between approximately 0.55 and 3$\times 10^{-11}$ergs cm$^{-2}$s$^{-1}$ over the 26.496 day orbit. This variability is echoed in the 2007-2008 dataset, however, the apparent modulation with the orbital phase present in the 2006-7 data is no longer evident.. The 2007-2008 RXTE dataset also shows the presence of three exceptionally large X-ray flaring points, reaching a peak flux on MJD 5436.96 of 7.2$\times 10^{-11}$ergs cm$^{-2}$s$^{-1}$ during a $\sim$500s integration window. This flare stands as the largest X-ray flare detected from this source, a factor of 2-3 larger than the previously measured flares. There are several other powerul X-ray flares occuring shortly after this flare in the RXTE data. It should be noted that the Swift data points which are temporally close to the RXTE flare points do not show evidence for such strong flaring episodes. However, the observation windows do not overlap and are separated by several hours. This variability is examined in further detail in Smith et al. (2008).

\subsection{Combined Datasets}

It is not clear from the data presented here that a consistent correlation exists between emission in the two bands. In figure 3 it can be seen that there are regions of significant TeV emission detected contemporaneously with X-ray flux which is above the average value ($\sim$1.5$\times 10^{-11}$ergs cm$^{-2}$s$^{-1}$), for example October 27-30 2006. However, there are instances of significant TeV detections with only marginally elevated X-ray emission present as seen in the observations taken between October 8-11, 2007. Although no obvious correlation is present between the two bands, it should be noted that the TeV sampling is relatively sparse and no defiinitive conclusions about a possible correlation can be made at this time. A formal correlation study of the entire dataset will be presented in \cite{Accb2008}.

Since no spectral information is available during strictly simultanous observations in both X-ray and TeV gamma rays, it is only possible to construct a quasi-contemporaneous SED with the current dataset (figure 4). In this figure is shown the VERITAS energy spectrum for LS I +61 303 which was derived only from the phases 0.5$\rightarrow$0.8 in 2006-7 (when the source was active in that season). This spectrum is plotted along with the EGRET spectrum for the source from \cite{Har2000}. For the X-ray spectrum, all data taken between orbital phases 0.5$\rightarrow$0.8 by RXTE was binned and a single spectrum was extracted. The shaded areas in figure 4 represent the ranges in SED possible by extrapolating the associated error bars in each band. It can be seen from this figure that (regardless of emission mechanism) is is unlikely that emission in all three bands can be fit by a single powerlaw.

\section{Summary}

VERITAS has detected LS I +61 303 as a variable source in the $>$300 GeV energy between 2006 and 2008. These observations show large changes in both detected flux maximum from year to year, as well as location in orbital phase where the largest emission takes place. It is clear that further, more comprehensive TeV observations must be performed before a claim of strict periodicity can be made in the TeV regime. The X-ray campaign performed by RXTE and Swift over the same time period shows similarly large variations in flux maxima and location of maxima along the orbit. While previously believed to be the most active in X-rays around phases 0.5-0.8, it is clear from the 2007-8 RXTE monitoring that strong X-ray activity is occuring at varying points of the orbital phase. The large flares detected in the RXTE data stand as the strongest X-ray activity detected from this source to date.

Comparison of the overlapping X-ray and TeV data do not show a clear correlation between emission states at this time; however, due to the lack of densely sample TeV data, a correlation cannot be ruled out at this time. A more formal statistical correlation study between the two bands, as well as additional data taken during VERITAS moonlight observations is presented in \cite{Accb2008}. The comparison of the derived X-ray, GeV, and TeV spectra show that it is unlikely that emission in all three bands can be described by a single powerlaw.  Although no definitive statement about ruling out or confirming possible scenarios to describe LS I +61 303 (i.e. microquasar or binary pulsar) can be made at this time, it is clear that further multi-wavelength campaigns such as the work described here may aid in modeling of the system.

\pagebreak

\begin{theacknowledgments}
This research is supported by grants from the U.S. Department of Energy, the U.S. National Science Foundation, and the Smithsonian Institution, by NSERC in Canada, by PPARC in the UK, and by Science Foundation Ireland.
\end{theacknowledgments}



\bibliographystyle{aipproc}   % if natbib is available

\begin{thebibliography}{1}


\bibitem{HC1981}Hutchings, J.D. and Crampton, D.,  \emph{PASP}, \textbf{93}, 486, (1981)
\bibitem{C2005} Casares, J. et al., \emph{MNRAS}, \textbf{360}, 1105, (2005)
\bibitem {Par1998}Paredes, J.M. et al., \emph{A$\&$A}, \textbf{335}, 539, (1998)
\bibitem{Greg2002}Gregory, P.C., \emph{ApJ}, \textbf{525}, 427, (2002)
\bibitem{L1997} Leahy, D.A. et al., \emph{ApJ}, \textbf{475}, 823, (1997)
\bibitem{Tay1996}Taylor, A.R. et al.,\emph{ A$\&$A}, \emph{305}, 817, (1996)
\bibitem{GR2001}Greiner, J. and Rau, A.,\emph{A$\&$A}, \textbf{375}, 145, (2001)
\bibitem{Har2000}Harrison, F.A. et al., \emph{ApJ}, \textbf{528}, 454, (2000)
\bibitem{M2004}Massi, M., \emph{A$\&$A}, \textbf{422}, 267, (2004).
\bibitem{Alb2006}Albert, J. et al, \emph{Science}, \textbf{312}, 1771, (2006) 
\bibitem{Acc2008}Acciari, V. et al., \emph{ApJ}, \textbf{679}, 1427, (2008)
\bibitem{W2002} Weekes, T. et al., \emph{Astropart Phys}., \textbf{17}, 221, (2002)
\bibitem{Gernot2007} Maier, G. et al., \emph{Proc. of  30$^{th}$ ICRC, Merida, MX}, (2007)
\bibitem{Swank2004} Swank, J.H., \emph{In Proc. of 185$^{th}$ American Astronomical Society Meeting} \textbf{64}, 1620, (1994)
\bibitem{Smith2008} Smith, A.W. et al., submitted to ApJ, astro-ph/0809.4254
\bibitem{G2004} Gehrels, N. et al., \emph{ApJ}, \textbf{611}, 1005, (2004)
\bibitem{H2007} Holder, J., Falcone, A., Morris, D., \emph{In Proc. 30$^{th}$ ICRC., Merida, MX}, (2007)
\bibitem{Accb2008} Acciari, V. et al., \emph{In Preparation}, (2008)

\end{thebibliography}



%%%%%%%%%%%%%%%%%%%%%%%%%%%%%%%%%%%%%%%%%%%
%% The following lines show an example how to produce a bibliography
%% without the help of the BibTeX program. This could be used instead
%% of the above.
%%%%%%%%%%%%%%%%%%%%%%%%%%%%%%%%%%%%%%%%%%%

\end{document}